\documentclass[pdf-a,balance,colorlinks,upint,subscriptcorrection,varvw,mathalfa=cal=boondoxo, spanish,french,vietnamese,russian,greek]{asmeconf}

\usepackage{rotating}
\usepackage{mwe}
\usepackage{graphbox}
\usepackage{tabularray}
\usepackage{float}
\usepackage[T1]{fontenc}
\usepackage[utf8]{inputenc}
\usepackage{lmodern}
\usepackage{graphicx}
\usepackage{tikz-cd}

\usepackage{fancyhdr}
\newcommand{\etal}{\textit{et al.}}

\begin{document}


\title{Prompting for Products: Investigating Design Space Exploration Strategies for Text-to-Image Generative Models}




\SetAuthors{%
	Leah Chong\affil{1}, 
        I-Ping Lo\affil{2}
	Jude Rayan\affil{3}
	Steven Dow\affil{3}, 
        Faez Ahmed\affil{1},
	Ioanna Lykourentzou\affil{2}\CorrespondingAuthor{i.lykourentzou@uu.nl} 
	}

\SetAffiliation{1}{Massachusetts Institute of Technology, Cambridge, MA}
\SetAffiliation{2}{Utrecht University, Utrecht, Netherlands}
\SetAffiliation{3}{University of California, San Diego, San Diego, CA}

\maketitle

\begin{abstract}
Text-to-image models are enabling efficient design space exploration, rapidly generating images from text prompts. However, many generative AI tools are imperfect for product design applications as they are not built for the goals and requirements of product design. The unclear link between text input and image output further complicates their application. This work empirically investigates design space exploration strategies that can successfully yield product images that are feasible, novel, and aesthetic - three common goals in product design. Specifically, users' actions within the global and local editing modes, including their time spent, prompt length, mono vs. multi-criteria prompts, and goal orientation of prompts, are analyzed. Key findings reveal the pivotal role of mono vs. multi-criteria and goal orientation of prompts in achieving specific design goals over time and prompt length. The study recommends prioritizing the use of multi-criteria prompts for feasibility and novelty during global editing, while favoring mono-criteria prompts for aesthetics during local editing. Overall, this paper underscores the nuanced relationship between the AI-driven text-to-image models and their effectiveness in product design, urging designers to carefully structure prompts during different editing modes to better meet the unique demands of product design.


\end{abstract}

\section{Introduction}
Rapid advancements in generative artificial intelligence (GenAI) have enabled the generation of novel and innovative content, such as texts and images, from simple text prompts. In product design applications, text-to-image models can produce images of designs from text prompts, enabling the exploration of multiple designs in shorter spans of time compared to the traditional method of manually rendering new designs. This functionality holds great potential for streamlining the iterative creative process in product design, particularly by facilitating design space exploration (DSE). 

While text-to-image GenAI can enable the rapid exploration of diverse product design concepts, most existing tools are not engineered to account for the multifaceted goals and requirements of product design, such as feasibility and aesthetics. For example, current GenAI tools can generate a large number of designs, many of which are infeasible \citep{feasible1,feasible2,feasible3}. Chong and Yang presented a list of 16 different design objectives that are prevalent in design research and practice \citep{chong}. This long -- and yet non-exhaustive -- list, underscores the complexity of parameters essential for designing a successful product. Unfortunately, current GenAI possesses AI's inherent vagueness in the relationship between the input (i.e., communicated goal) and the generated output (i.e., images of designs), a property that renders GenAI tools insufficient for generating reliable product designs. For example, when the text prompt is “design of a mug that is ergonomic, sleek, and modern”, it is unclear how the GenAI understands and maps the meaning of “ergonomic”, “sleek”, and “modern” onto the generated images. While one way to address this problem is to develop new models specifically trained for product design, current models possess a creative advantage, given the vast range of available image datasets compared to design-specific datasets like computer-aided design files. Therefore, this work aims to understand how off-the-shelf, promising yet imperfect text-to-image GenAI tools can be used to explore and refine product designs that are novel, aesthetically pleasing, and feasible.

This work conducts a controlled human subject experiment in which participants are asked to design bikes that are feasible, novel, and aesthetic at the same time using Stable Diffusion 1.5 on an online platform called Leonardo.AI. At the time of running the experiment, Leonardo.AI was one of the few interactive tools that allowed the participants to easily use text-to-image generative models, such as various versions of Stable Diffusion, without the need to run custom Python scripts. The evaluations of the feasibility, novelty, and aesthetics of the generated designs are collected via crowdsourcing. The relationship between the participants' DSE strategies when using Stable Diffusion and the evaluation scores of their generated designs are analyzed. Key results from this study show that the goal orientation and the number of goals targeted by the prompts are more closely correlated with the design evaluation scores than the time users spend editing globally versus locally and their length of prompts. During early, broader exploration stages, multi-criteria prompts with feasibility and/or novelty goal orientation are found to be effective. Later in more refinement-focused stages, mono-criteria, aesthetics-oriented prompts are suggested to be used. 

The rest of this paper is organized as follows. The next section provides a review of the background literature on DSE and text-to-image GenAI, identifying the research gap. Then, the purpose and the impact of this work are discussed, followed by the Method section, illustrating the study design, including descriptions of the task, participants, and experimental procedure. Then, the results are presented and discussed, including a discussion of the limitations of this work. The paper concludes with a summary of the main findings and its implications for design research and practice.

\section{Background}
\subsection{Design Space Exploration}

DSE is a crucial step in the product design process, during which designers explore a wide range of potential designs (divergence) and select and refine a fewer selection of designs (convergence) \citep{cross, georgiades2019adopt}. Divergence is important for successful design as it expands designers' creativity and increases the novelty and quality of their designs \citep{hilliges2007designing}. It aims to prevent designers from limiting themselves to one or few viable solutions too early by encouraging the formulation of a variety of potential solutions. Along with divergence, convergence is an equally important aspect of DSE. Once designers have explored enough, they must evaluate, select, and refine the final solution(s) based on various design requirements, goals, and preferences. 

During DSE, designers engage in divergent thinking through various methods like problem reframing \citep{foster2021design, schon1987educating} and analogies \citep{mose2017understanding}. Generation and consideration of a large number of design options can be encouraged through the manipulation of a variety of characteristics, such as flexibility and imagery \citep{guilford1956structure}. This process not only allows designers to maximize their creative output, but also provides an opportunity to gain more insights into the problem and the design space. Prior research has attempted to find effective inspirations and methods to assist designers' divergent thinking using non-AI-based methods. For example, Thinklets \citep{briggs2001thinklets} is a creativity support tool that guides designers to think of ideas from multiple angles through open-ended questions. Ideation Decks \citep{golembewski2010ideation} is another example, a set of cards that prompts designers to think about their design space from different angles. While these tools have shown some effectiveness, text-to-image GenAI can enable a much more efficient DSE and significantly reduce designers' cognitive load by quickly generating multiple designs from text prompts. However, adopting this methodology means that designers must sacrifice some level of control in the design generation process (between text prompt and generated image). 

Convergence is also a crucial aspect of DSE, during which designers make selections and/or mark preferences for certain aspects of generated designs \citep{mose2017understanding}. Often informed by data, designers choose specific design directions and make refinements to the designs as the most important dimensions of the problem space come into focus. Tools to support generative design exploration apply convergence methods in different ways. For example, the workflow in Dream Lens \citep{matejka2018dream} starts with the user defining the problem space. Then, the constraints and requirements from the resulting definition are used by the algorithm to generate a design. Additionally, GANCollage \citep{wan2023gancollage} updates its backend with the "user selection" every time the user requests "similar images" to accomplish the objective of choosing one final image. For effective design convergence, it is crucial to understand various user interactions during image exploration that could drive this process. Leonardo.AI, the tool used in this study, also includes various features that aim to facilitate design convergence, which will be explored in this work.

\subsection{Image Generative AI}
With the recent advances in GenAI, there is great potential for these tools to effectively support creative processes. GenAI is a rapidly evolving field that involves the creation of algorithms and models capable of generating novel content in various domains, such as images, text, and music. Its primary goal is to imitate the intricate creative process by leveraging existing datasets to identify underlying patterns and yield outputs that closely resemble the characteristics of the training examples. Since 2020, discussions of various applications of GenAI, such as human resources, literature, and art, have emerged. Specifically in product design, the potential of image GenAI as a tool for the early stages of the design process has been explored in some recent design literature \citep{lee2023impact}. There are primarily two types of models employed for image GenAI: Generative Adversarial Networks (GANs) and diffusion models. 

GANs were introduced by Goodfellow in 2014 in the field of machine learning (ML). GANs are built using a pair of neural networks: a generator and a discriminator, operating on the principle that one network's gain is another network's loss. The generator is trained to generate new data samples, while the discriminator determines whether these samples are real or generated. Training continues until the discriminator's performance is above a certain threshold. Over the years, GANs have undergone significant refinement, incorporating methods such as injecting noise into the generator's input \citep{salimans2016improved}, employing diverse loss functions \citep{mao2017least}, and applying regularization methods \citep{miyato2018spectral}, to promote the diversity of the generated data and improve the overall quality of the model outputs. With this refinement, the practical applications of GANs have expanded, serving as an effective model for image-to-image GenAI for DSE by rapidly creating a large number of possible designs \citep{chen2021padgan, li2021product}.

The diffusion model was introduced by Sohl-Dickstein, et al. in 2015 as an alternative paradigm for GenAI \citep{nichol2021improved}. The diffusion model works by adding small random noise to the training data over multiple steps to produce a sequence of samples, then learning to recover the data by reversing this process. The performance of the model has been advanced continuously, giving rise to a flow-based generative model employing invertible transformations \citep{dinh2016density},  a continuous-time diffusion process called the Free-form Jacobian of Reversible Dynamics (FFJORD) model capable of generating high-quality samples with efficient inference  \citep{grathwohl2018ffjord}, and a new architecture that combined flow-based models with invertible 1x1 convolutions \citep{kingma2018glow}. Ho, et al. further improved flow-based generative models, enhancing the quality and diversity of generated samples \citep{ho2019flow++}. Most of the current, widely-used text-to-image GenAI tools like DALL·E 2, Stable Diffusion, and Midjourney are founded on diffusion models.

Diffusion models offer unique advantages compared to GANs. They guarantee a more fine-grained control over the generated images, permit data quality and diversity manipulation \citep{giambi2023conditioning}, and avoid mode collapse via a stable training process. A paper by OpenAI researchers \citep{dhariwal2021diffusion} has indicated that diffusion models can achieve image sample quality superior to the GAN models. However, some main drawbacks of diffusion models are that they require longer training times and are computationally expensive because of the model's inherent complexity and the sequential nature of the diffusion process. This work uses Leonardo.AI, an online GenAI platform that provides an interface to work with various text-to-image diffusion models. Leonardo.AI is a particularly appropriate tool for this work as it offers global and local editing modes that are useful for investigating the users' DSE process. At the time of the experiment, Leonardo.AI was one of the few, if any, online text-to-image GenAI platforms that allowed the users to seamlessly transition between global and local editing modes.

\subsection{Research Gap}
Despite the promises of text-to-image GenAI in aiding the engineering design process, many current tools are imperfect tools for product design applications because of the various design goals and requirements, as well as AI's inherent vagueness in the relationship between the input (i.e., communicated goal) and the generated output. Therefore, there is an open question on how to use these promising yet imperfect GenAI tools for DSE and yield desirable designs. Only when this question is answered can people successfully utilize text-to-image GenAI for product design. 

\section{Research Aims and Significance}

\begin{figure*}[h!]
    \centering
    \includegraphics[width=0.8\linewidth]{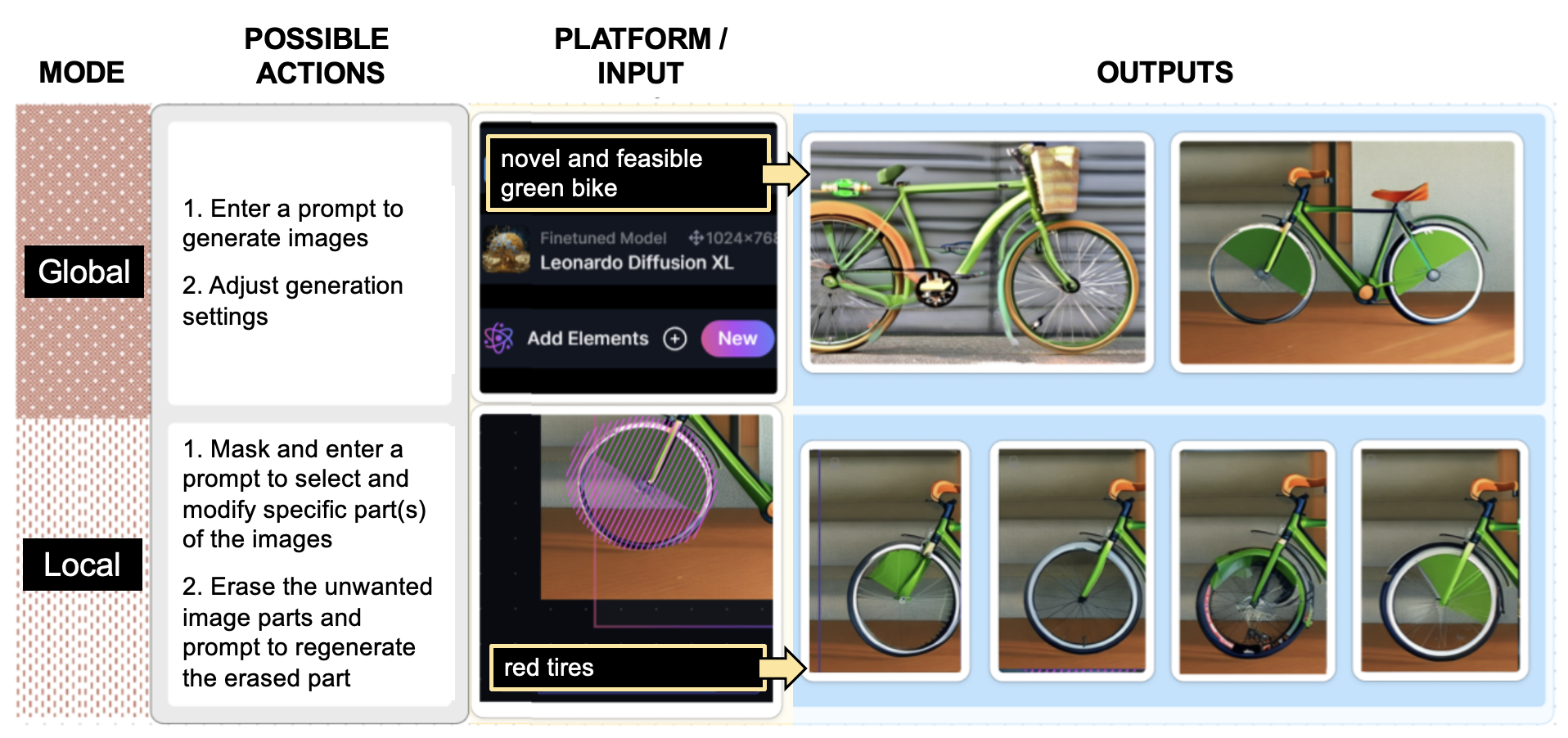}
    \caption{Global and Local Editing Modes in Leonardo.AI and Their Example Input and Output. The global editing mode is primarily for generating entire images by entering text prompts, while the local editing mode is for more detailed refinement of selected images using features like prompting, masking, and erasing.}
    \label{fig:leonardo}
\end{figure*}

The purpose of this work is to close this gap in knowledge by investigating different DSE strategies using text-to-image GenAI and their impact on the generated outcome, specifically feasibility, novelty, and aesthetics. The strategies are observed through the users' interaction with and during the global and local editing modes on Leonardo.AI. Figure \ref{fig:leonardo} displays what global and local editing modes entail in Leonardo.AI. The global editing mode is primarily for generating entire images by entering text prompts, while the local editing mode is for more detailed refinement of selected images using features like prompting, masking, and erasing. The research question of this work is:
\begin{quote}
How do users' design space exploration strategies when using text-to-image generative AI influence the feasibility, novelty, and aesthetics of the generated product designs?
\end{quote}

Specifically, we want to understand whether and how the time users spend and the characteristics of prompts used in global and local editing modes have significant impact on the feasibility, novelty, and aesthetics of the generated outcomes. It is hypothesized that the more time spent in the global editing mode exploring the design space, the better the ratings of the generated outcome. Additionally, it is expected that the more prompting is focused on a goal, the more likely the rating for that goal will be higher. This work is expected to contribute to the design research community by suggesting what DSE strategies are effective when using GenAI for product design, specifically for designing feasible, novel, and asethetic products.

\section{Method}
A human subject experiment is designed and conducted to examine how users leverage global and local editing modes in Leonardo.AI, an online GenAI platform, to explore and create feasible, novel, and aesthetic designs. The collected data are analyzed to find DSE strategies that yield outcomes that successfully meet the design goals. The feasibility, novelty, and aesthetic ratings for generated designs are collected via crowd-sourced evaluations. 

\subsection{Human Subject Experiment}


The entire experimental process is recorded and transcribed, facilitating the observation and analyses of user actions. During the experiment, the participants interact with Stable Diffusion via the Leonardo.AI platform to complete a design task. Leonardo.AI offers a wide range of features, as shown in Fig. \ref{fig:leonardo}.

\subsubsection{Participants}
Total 15 participants are recruited and have completed the experiment. They range over ages between 25 and 33 years and vary in their level of experience with GenAI tools. Regarding the level of experience with text-to-image GenAI, nine participants are first time users, three are somewhat inexperienced, two are neither inexperienced or experienced, and one is somewhat experienced. Six of the participants are male, and nine are female. Their educational backgrounds range from bachelors to doctorate degrees. Additionally, all participants demonstrate proficiency in English above level C1 and use English on a day-to-day basis.

A Google form for recruitment is employed to gather essential participant information, such as age, email address, education, and level of experience with GenAI tools. It is also ensured that they viewed the tutorial video on Leonardo.AI and that their consent is obtained for participation and recording. Every participant are compensated with a 20 euro voucher for their participation in the experiment. 

\subsubsection{Experimental Task}
The participants are asked to create feasible, novel, and aesthetically-pleasing bike design(s) using Stable Diffusion on Leonardo.AI in 30 minutes. Feasibility is a crucial design goal as it assures that the tools align with real-world constraints. It examines whether the AI-generated images can realistically be implemented within the context of a product. This is essential to avoid unattainable designs. This work also considers novelty to ensure that users harness the potential of text-to-image GenAI in producing creative designs \citep{Mukherjee2023}. Finally, aesthetics goal is demanded in this work as well it is a significant component in design as a major factor of product popularity \citep{Bloch2003}. 

To provide a clear design context, the participants are given the "bike product designer" persona for a company named "22-Century Bike", as well as instructions to submit any number of bike designs. Throughout the experiment, they are encouraged to think-aloud to collect data on their thoughts and decision-making processes. The following task description is provided to the participants: 
\begin{quote}
You work for the company 22-century Bike as a bike product designer. Your job is to make one or more new bike designs. Your new bike design(s) should be feasible to manufacture (no triangle wheels!) and as unique/novel and aesthetically pleasing as possible. You can submit multiple designs. 

Remember to think aloud!
The time will be about 30 mins.
\end{quote}

\subsubsection{Procedure}
At the start of the experiment, the participants are given a comprehensive pre-task tutorial. This tutorial aims to familiarize the participants with Leonardo.AI, explain the experimental task, and ensure that they possess adequate knowledge about bikes. First, to facilitate the participants' usage of Leonardo.AI's features, all features are demonstrated to them within the tool itself. Additionally, the participants are provided with a wiki page containing information about bike parts and types. Then, they are asked to complete a practice task on creating a vase for tulips using Leonardo.AI, giving the participants a chance to engage with the software and get their questions answered.

After the pre-task tutorial session, the main task begins. The participants are provided with the task description and are given 30 minutes to complete the task. Once the task is completed, the participants undergo a post-experiment interview and a brief questionnaire. The three interview questions are:

\begin{itemize}
    \item Can you briefly describe your experience using Leonado.AI? What did you like about it? What did you dislike about it?
    \item Did you encounter any difficulties using the tool? If so, can you describe what they were?
    \item Are there any additional features or functionalities you would like to see added to the tool to improve?
\end{itemize}
These additional measures aim to gather qualitative data to understand the quantitative results further. Then, the questions in the post-experiment questionnaire ask the participants to report how easy it was to use each feature of the tool, how important they think each feature of the tool is, and how feasible, novel, and aesthetic they think their final designs are. These questions are answered in a 5-point Likert scale. 

\subsection{Crowd-Sourced Design Evaluations}
\begin{figure}[h!]
    \centering
    \includegraphics[width=1.05\linewidth]{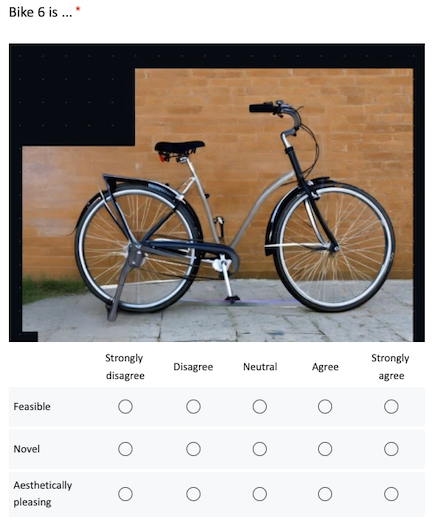}
    \caption{Example Problem in the Crowd-Sourced Image Evaluations}
    \label{fig:crowd}
\end{figure}

The final set of 18 images is submitted by the 15 participants (one image each by 12 participants and two images each by three). Crowd-sourced evaluations are conducted using Google Forms to assess the feasibility, novelty, and aesthetics of these designs. The evaluation questionnaire asked raters to evaluate each bike image on its feasibility, novelty, and aesthetics based on a 5-point Likert scale ranging from 'Strongly Disagree' to 'Strongly Agree'. Some example questions are shown in Fig. \ref{fig:crowd}.

Raters are recruited from the Prolific platform, which is a platform for researchers to recruit and manage participants from a large, global participant pool for online studies. In total, 10 raters evaluated the images, allowing for a broader range of perspectives and opinions to be considered. 

\section{Results}
In this work, DSE strategies are observed by the participants' actions the global and local editing modes in Leonardo.AI. Within each editing mode, the participants' actions are mostly done via text prompting. Therefore, this work examines the participants' prompting action characteristics to gain insight into their exploration strategies. Along with the participants' exploration strategies, this section also presents the correlation results between these strategies and the feasibility, novelty, and aesthetics ratings of the generated outcomes. 

Three major action characteristics in the global and local editing modes are examined: length, goal orientation, and multi vs. mono-criteria of prompt. These three characteristics have been selected because they are frequently discussed in the prompt engineering literature. For instance, Xie, et al. performed a log analysis from participants from a text-to-image and found a correlation between the length of the prompt and the quality of the generated image \citep{xie2023prompt}. From a prompt construction perspective, studies have discussed about mono-criteria prompts having higher accuracy than multi-criteria prompts \citep{tan2020progressive, wei2022chain}, meaning that targeting multiple design objectives in a single prompt might not accurately express the designer's intentions. Finally, PromptMagician is a tool that suggests keywords to be added to the text prompt to enhance the alignment of the generated images with the intended vision of the creator \citep{feng2023promptmagician}. 

It is important to note that three out of the 15 participants submitted two bike designs and none more than two. For these three participants, the average data of their two bike creation processes are used throughout the analyses. Therefore, in this work, one data point consistently represents each participant. 

Before conducting each statistical analysis presented in this section, normality of the included data is tested via Shapiro-Wilk test. If the data are normal, two-sample t-test and Pearson's correlation test are conducted for two sample comparison and correlation analyses respectively. If the data are not normal, Wilcoxon signed-rank test and Spearman's rho test are conducted. 

\subsection{Global vs. Local Editing}
\begin{figure*}[h!]
\centering
\includegraphics[width=0.65\linewidth]{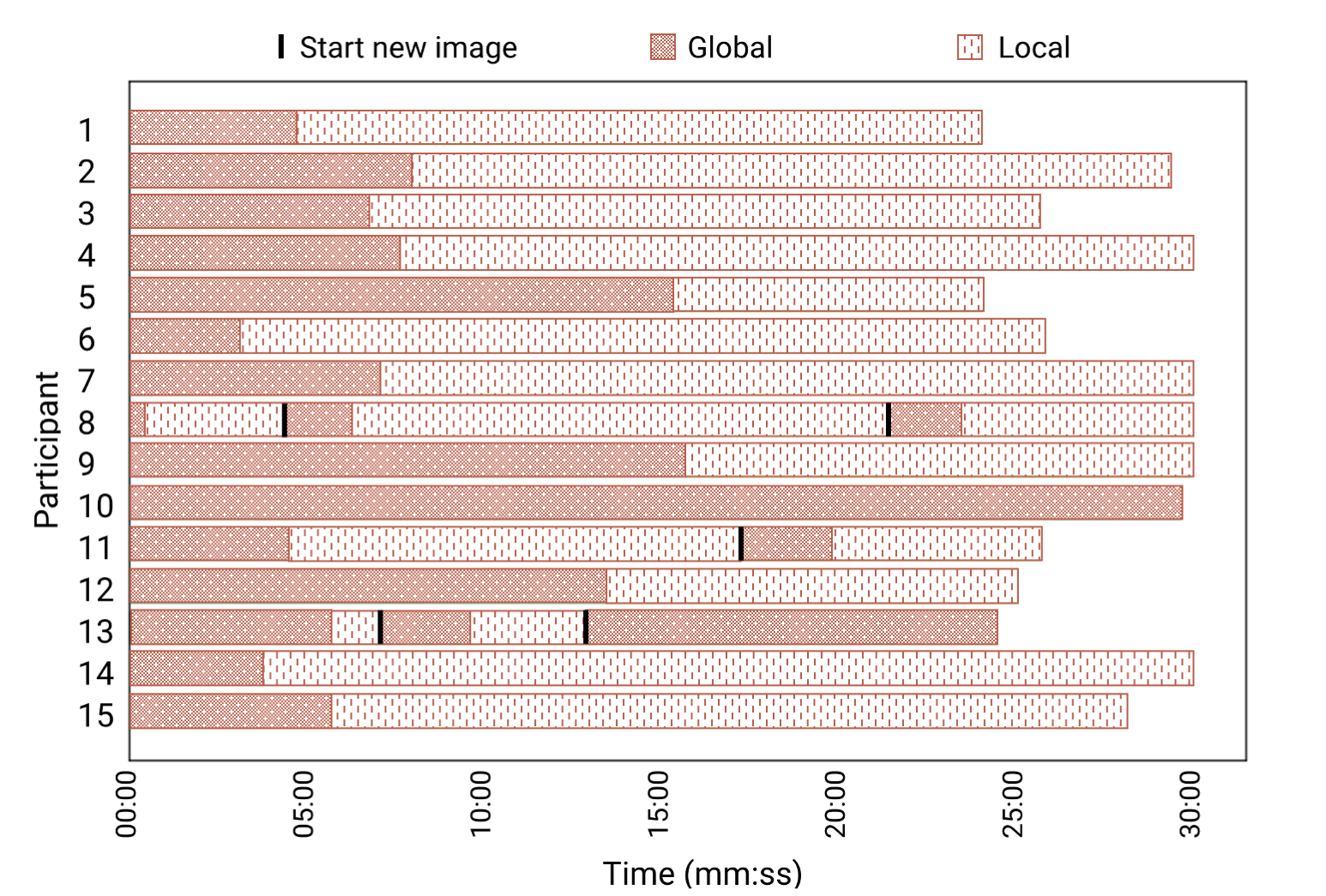}
\caption{The Distribution of Time Spent in Global and Local Editing Modes. On average, the participants spent about 38.5\% of their time in the global editing mode, and the other 61.5\% in the local editing mode.}
\label{fig:GL_time}
\end{figure*}

One notable aspect of users' exploration strategy with respect to Leonardo.AI is how they choose to allocate their time between the global and local editing modes. As shown in Fig. \ref{fig:leonardo}, the global and local editing modes are available on Leonardo.AI for the participants to either change the entirety of the AI-generated image using text prompts or to change only a part of this image using various features like masking and erasing, respectively. The visualization of each participant's time distribution among global and local editing modes is shown in Fig. \ref{fig:GL_time}.

Most participants (14 out of 15) engage in both global and local editing, while only one user exclusively focused on global editing. It is observed that participants all start in the global editing mode, and after 4.0 prompts in the global editing mode on average, are inclined to select an image for local refinement. Furthermore, it was observed that the participants who entered the local editing mode never reverted back to the global editing mode, unless they are starting a new design.

On average, the participants allocate about 38.5\% of their time to global editing, concentrating on general modifications, and the remaining time (61.5\%) to local editing, refining specific details to meet the design goals. To investigate the split between global and local editing modes further, the number of prompts the participants enter in the two modes are observed. About 36.7\% of the prompts (on average 4.0 prompts) are entered in the global editing mode and the rest (63.3\%) in the local mode, a fairly consistent result with the time split result. Overall, the participants spend more time in the local editing mode than in the global editing mode, though without statistical significance (Wilcoxon signed-rank test, p=0.2 for time and p=0.1 for number of prompts). 


\subsection{Prompt Length}
\begin{figure*}[h!]
\centering
\includegraphics[width=0.65\linewidth]{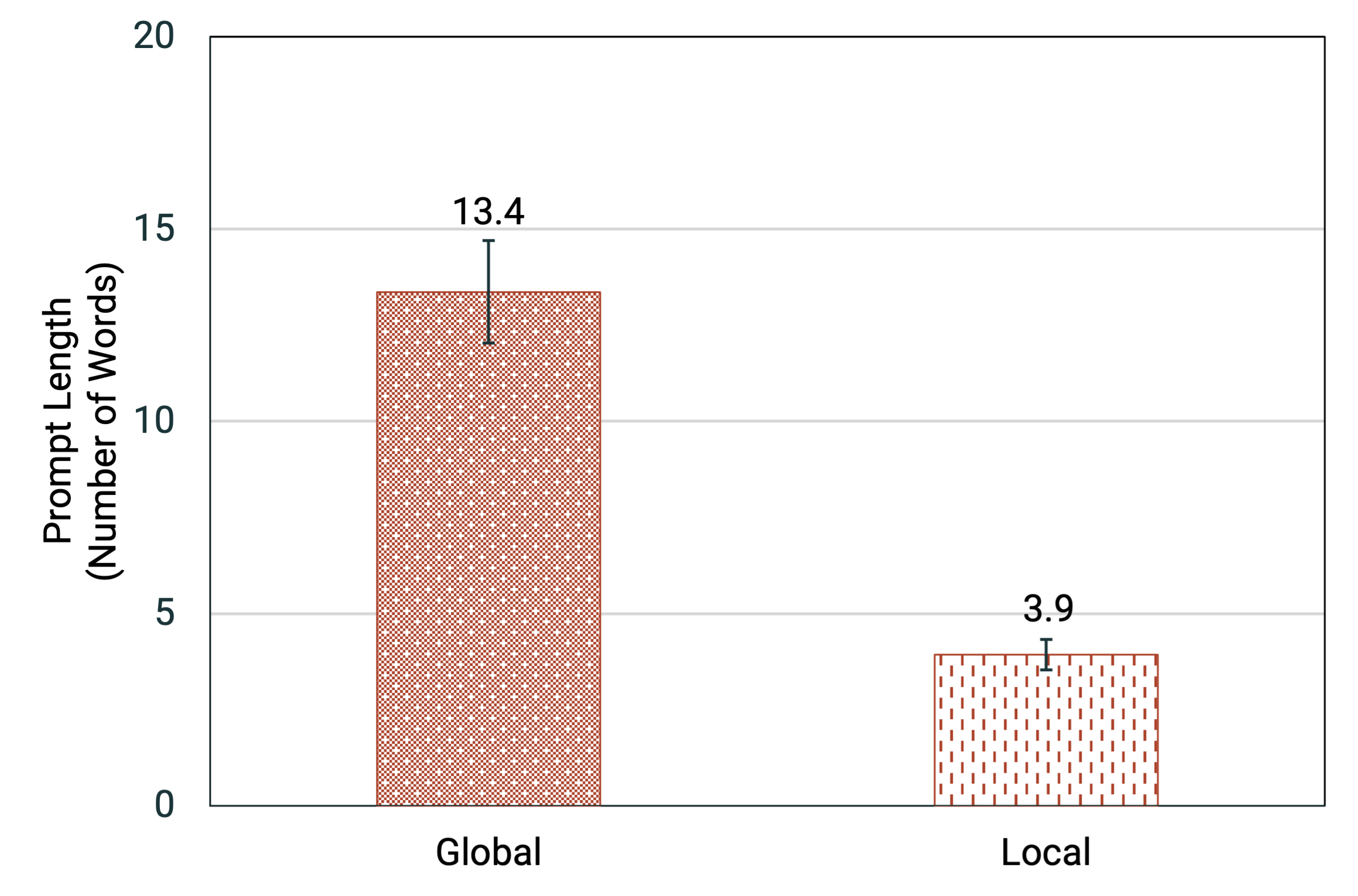}
\caption{Average Prompt Length. The participants use significantly longer prompts in the global editing mode compared to the local editing mode.}
\label{fig:length}
\end{figure*}

The participants' actions via prompting in the global and local editing modes are explored to extract their exploration strategies. The first action characteristic is the length of prompts. The average length of prompts are demonstrated in Fig. \ref{fig:length}. On average, the participants' prompts are about 7.7 words long, and the prompts they use in the global editing mode (13.4 words long on average) tend to be longer than those in the local editing mode (3.9 words long on average) (t-test, p=1.1e-3). The length of the longest prompt is 47 words, and the shortest has only one word.

\subsection{Mono vs. Multi-Criteria Prompts}
Given that the participants are asked to design for three different goals simultaneously, it is important to examine how they orient their prompts for the goals in the global and local editing modes. For this purpose, every prompt by the participants is labeled by the authors as feasibility, novelty, and/or aesthetics-oriented. Prior to labeling, the three authors agree on the descriptions of the three orientations of prompts shown in Table \ref{tab1}, which are used to guide the labeling process.

\begin{table*}[h!]
\caption{\label{tab1}Descriptions of Feasibility, Novelty, and Aesthetics-Oriented Prompts}
\centering
\begin{tabular}{l p{12cm}}
\hline
Prompt Orientation & Description \\
\hline
Feasibility & Contains explicit words related to manufacturing, addition, or modifications of bike parts, Is related to the usability (whether it successfully functions/can be used) and/or manufacturability \\
Novelty & Contains explicit words related to novelty or uniqueness, Prompting for any bike parts that are not in traditional bikes \\
Aesthetics & Contains any words related to the look/visual feel/dimensions of the bike \\
\hline
\end{tabular}
\end{table*}

\begin{figure*}[h!] 
\centering
\includegraphics[width=0.65\linewidth]{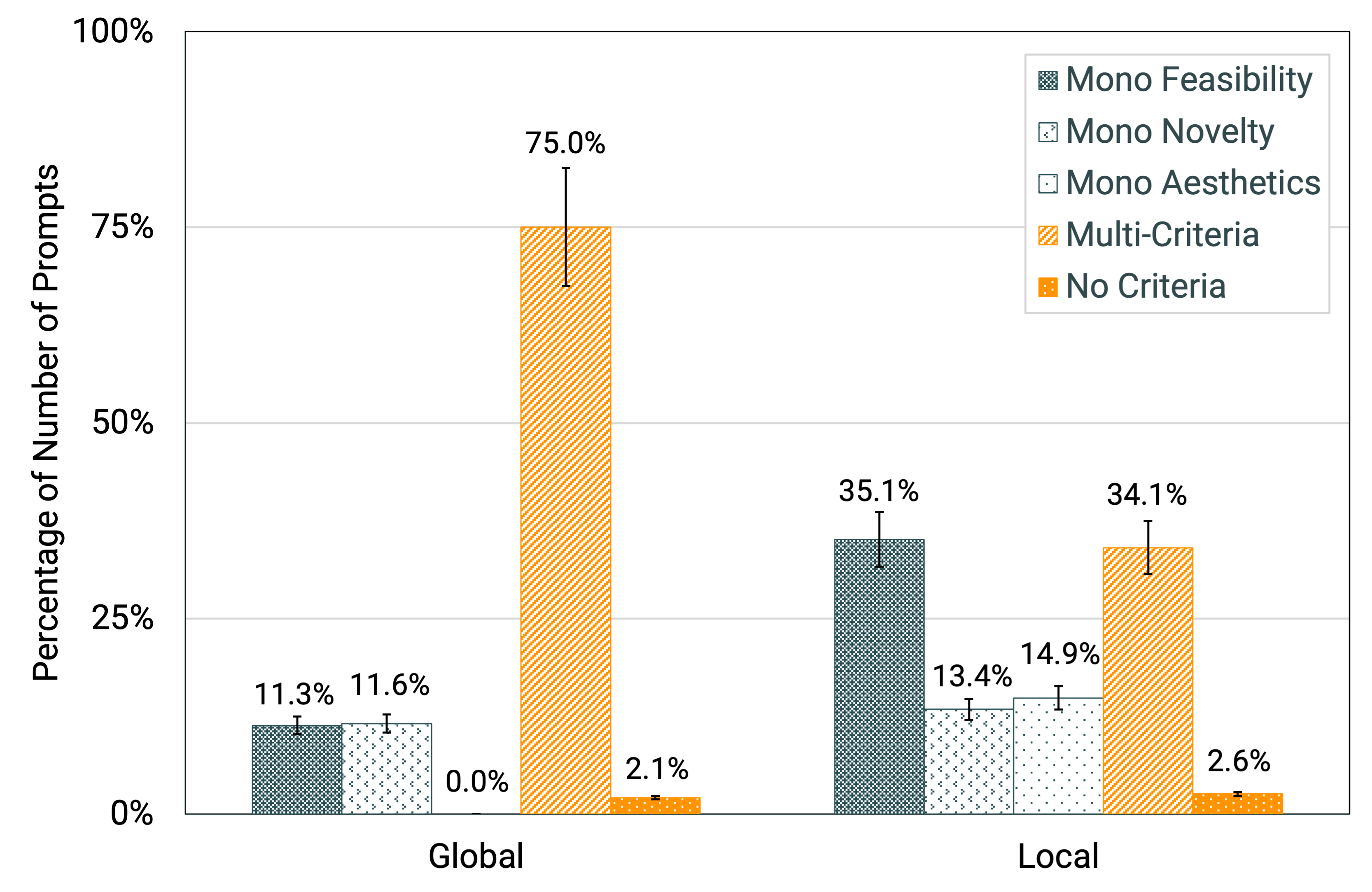}
\caption{Percentage of Mono-Criteria (Feasibility, Novelty, and Aesthetics-Oriented), Multi-Criteria, and No-Criteria Prompts. The participants use much more multi-criteria prompts in the global editing mode than in the local editing mode. Most of the mono-criteria prompts in the global editing mode are feasibility or novelty-oriented, while those in the local editing mode are mostly feasibility-oriented with some being novelty and aesthetics-oriented.}
\label{fig:monomulti}
\end{figure*}

The authors label the prompts separately according to the descriptions in Table \ref{tab1}, and the final labels are determined based on the majority agreement. Multi-criteria labeling is allowed. For example, if a prompt is relevant to both feasibility and novelty, it will be labeled as a both feasibility and novelty-oriented prompt. When there are negative prompts (prompts to eliminate components from an image), they are considered together with the main prompts. For instance, if the main prompt is primarily feasibility-oriented and the negative prompt is aesthetics-oriented, these prompts are considered as a single prompt that is both feasibility and aesthetics-oriented.

Many participants employ prompts that are oriented towards multiple design goals (i.e. multi-criteria prompts), such as "square tire bike with bottle cage and red chain ring" which is both novelty and aesthetics-oriented, while others concentrated on prompting for a single goal (i.e., mono-criteria prompts), such as "a bike with special design" which is only novelty-oriented. Therefore, the average percentage split of multi- vs. mono-criteria prompts in the global and local editing modes are observed to understand the participants' exploration strategy. Figure \ref{fig:monomulti} shows the results.

Overall, 52.0\% of the prompts are mono-criteria, and the rest (48.0\%) are multi-criteria, therefore showing a relatively equal split (t-test, p=0.7). This overall split is most likely because of the contrasting split in the global and local editing modes. In the global editing mode, there are more, though not statistically significant, multi-criteria prompts (75.0\%) than mono-criteria prompts (22.9\% = 11.3\% + 11.6\% + 0.0\%) (Wilcoxon signed-rank test, p=0.07), while in the local editing mode, there are more, though not statistically significant, percentage of mono-criteria prompts (63.4\% = 35.1\% + 13.4\% + 14.9\%) (Wilcoxon signed-rank test, p=0.1) than multi-criteria prompts (34.1\%). Therefore, the results show a tendency among the participants to tackle multiple design goals at once in the global editing mode, while taking one goal at at time in the local editing mode.

\subsection{Goal Orientation of Prompts}

\begin{figure*}[h!]
\centering
\includegraphics[width=0.65\linewidth]{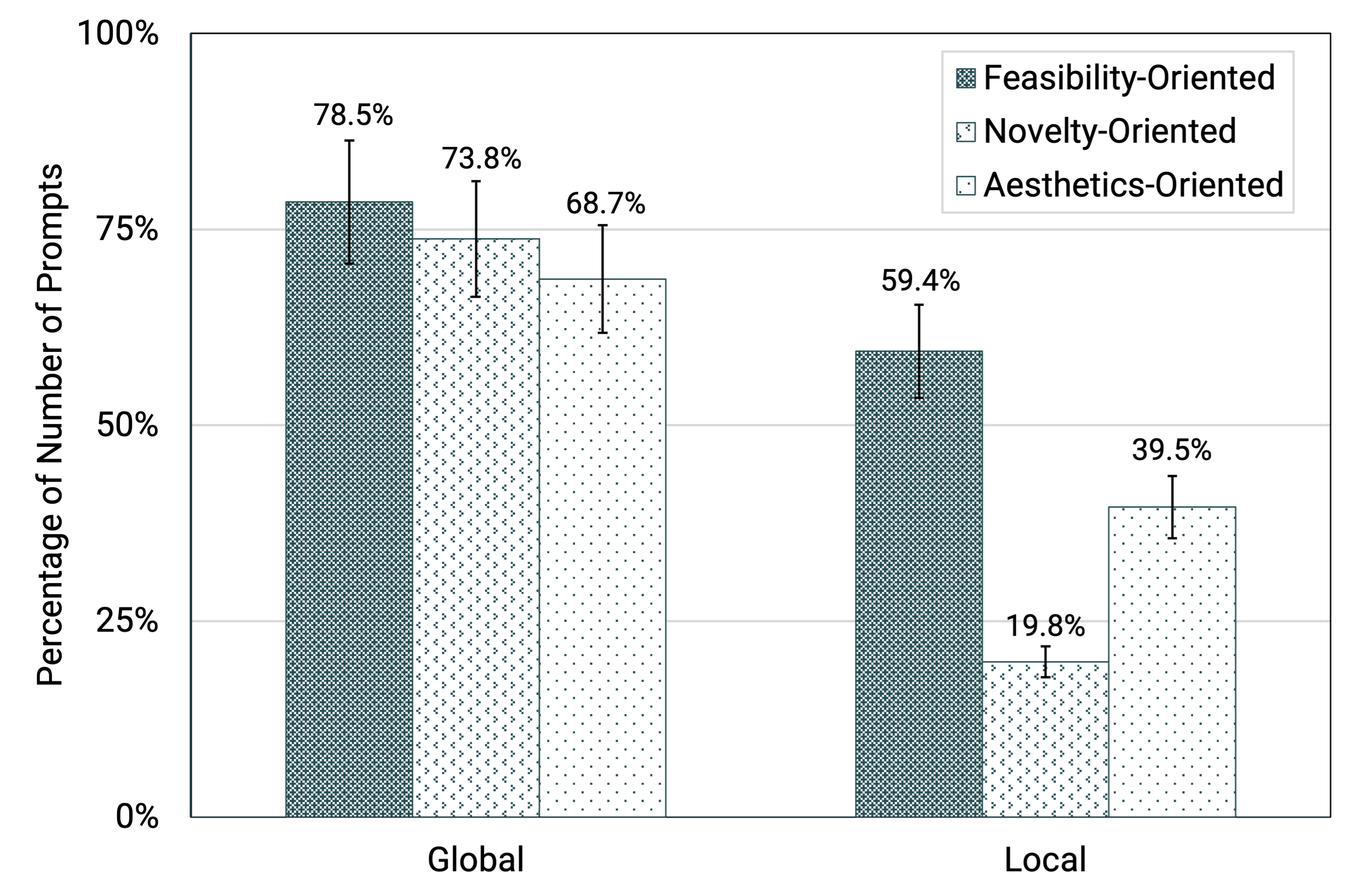}
\caption{Percentage of Feasibility, Novelty, and Aesthetics-Oriented Prompts. The results shown here include both mono and multi-criteria prompts, therefore not adding up to 100\%. While targeting all three goals often in the global editing mode, the participants tend to focus more on feasibility and aesthetics than novelty in the local editing mode.}
\label{fig:orientation}
\end{figure*}

Figure \ref{fig:orientation} shows the percentage split of feasibility, novelty, and aesthetics-oriented prompts. It is important to note that both mono and multi-criteria prompts are considered in these results; for example, the percentage of feasibility-oriented prompts includes the mono-criteria ones that are only feasibility-oriented, as well as the multi-criteria ones that target feasibility along with other goals. 

In the global editing mode, all three goals are targeted often in the participants' prompts without any statistical difference between the percentages of these orientations (78.5\% feasibility-oriented, 73.8\% novelty-oriented, and 68.7\% aesthetics-oriented). In contrast, in the local editing mode, the participants use a much higher percentage of feasibility-oriented prompts (59.4\%) than novelty-oriented prompts (19.8\%) (Wilcoxon signed-rank test, p=0.02). Comparing the results in the global and local editing modes, it is observed that the prompts in the global editing mode often target all three goals, while the prompts in the local editing mode are more likely to target feasibility over the other two goals.

\subsection{Correlation with the Design Outcome}
Correlation analyses between the observations above and the feasibility, novelty, and aesthetic ratings of the generated images are conducted to identify the exploration strategies that yield desirable design outcomes. The ratings are determined by the crowd-sourced evaluations described in the Method section. 

\subsubsection{Global vs. Local Editing and Prompt Length}
The correlations of the time the participants spent and their prompt length in the global and local editing modes with the design outcome are examined. No statistically significant correlations are found, meaning that neither how much time is spent nor how many prompts are used in the global versus local editing modes are related to any ratings of the design outcome.

\subsubsection{Mono vs. Multi-Criteria and Goal Orientation of Prompts}
\begin{figure*}[h!]
\centering
\includegraphics[width=0.65\linewidth]{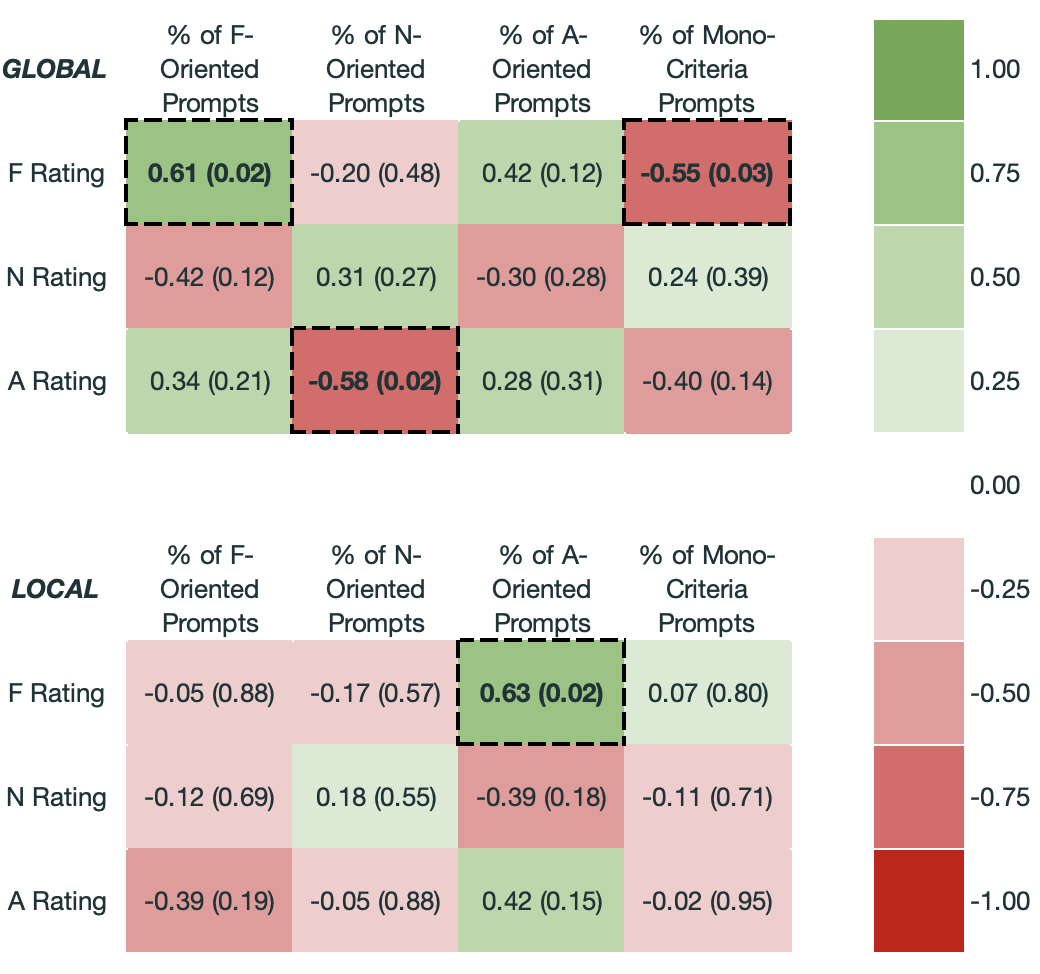}
\caption{Correlation Between the Prompting Characteristics (Mono vs. Multi and Goal Orientation) in Global and Local Editing Modes and the Crowd-Sourced Design Ratings (Feasibility, Novelty, and Aesthetics). The values are indicated as [correlation coefficient]([p-value]). The results bolded with dashed lines are statistically significant at 5\%.}
\label{fig:prompt_corr}
\end{figure*}

The participants' exploration strategy is also studied via two of their prompting characteristics in the global and local editing modes: mono versus multi-criteria  and goal orientation of prompts. The correlations between these characteristics and the feasibility, novelty, and aesthetic ratings of the bikes are computed, as demonstrated in Fig. \ref{fig:prompt_corr}. 

During the global editing mode, the percentage of feasibility-oriented prompts is significantly correlated to the feasibility rating of the generated image (Spearman's rho test, rho=0.6, p=0.02). This means that using more feasibility-oriented prompts during the global editing mode can significantly boost feasibility ratings. Therefore, if you want to enhance novelty in your prompts, it would be beneficial to use more prompts specifically targeting novelty-related aspects. Such correlations are not shown between novelty and aesthetics-orientated prompts and their corresponding ratings (Spearman's rho test, rho=0.3, p=0.3 for both). 

Interestingly, however, the percentage of novelty-oriented prompts in the global editing mode is negatively correlated to the aesthetics rating (Spearman's rho test, rho=-0.6, p=0.02). When more novelty-oriented prompts are used in the global editing mode, the generated design are less aesthetically-pleasing. 

Finally, the percentage split between mono and multi-criteria prompts in the global editing mode is found to be negatively correlated to the feasibility rating (Spearman's rho test, rho=-0.5, p=0.03), as well as the average rating (Spearman's rho test, rho=-0.6, p=0.01). More multi-criteria prompts and less mono-criteria prompts can help increase the feasibility and average ratings of the design outcome.

In the local editing mode, the only significant correlation is found between the percentage of aesthetics-oriented prompts and the feasibility rating of the outcome. This correlation is positive, meaning that the more aesthetics-oriented prompts are used in the local editing mode, the higher the feasibility rating of the generated image is (Spearman's rho test, rho=0.6, p=0.02).

The final result to note is the general positive relationship between feasibility and aesthetics, while they both show a negative relationship with novelty. The results regarding goal orientation of prompts in Fig. \ref{fig:prompt_corr} mostly demonstrate positive correlation coefficients between feasibility and aesthetics. Novelty has negative correlation coefficients with feasibility and aesthetics. This result is supported by the positive correlation between the percentage of feasibility-oriented and aesthetics-oriented prompts found in the Goal Orientation of Prompts section.

\section{Discussion} 
The purpose of this work is to address the research question: 
\begin{quote}
    How do users' design exploration strategies when using image generation AI tools influence the feasibility, novelty, and aesthetics of the generated outcomes?
\end{quote}
This section discusses the answers to this question with the results found in this work, as well as their implications for design research and the use of text-to-image GenAI in product design. Then, the limitations of this work and the areas for future research are discussed.

The results in this work first show that people commonly employ a combination of global and local editing with a significant focus on local refinements, spending approximately 61.5\% of their time locally editing the images. Once in the local editing mode, they do not return to the global editing mode. This behavior may be because of the complexity of the steps on Leonardo.AI to return to the global editing mode, especially given the time constraint. It could also be a reflection of "design fixation" \citep{bib10, bib11}, which describes people's tendency to overly focus on their initial ideas without considering diverse set of ideas. 

Despite observing similar human behaviors from prior works, this work demonstrates no relationship between the time spent editing the image globally and locally and the feasibility, novelty, and aesthetics ratings of the final design. At a glance, this is surprising in light of some prior findings about the positive impact of more time spent exploring the design space in the design process before converging to a solution \citep{kudrowitz2013getting}. However, it is important to note that the global and editing modes in this study are not exact reflections of divergence and convergence of ideas. Considering the features of text-to-image GenAI tools, such as Leonardo.AI, the users can easily fixate on a design even in the global editing mode by inserting conceptually similar prompts rather than exploring different design options with conceptually far prompts. Therefore, the most precise way to interpret the result is that the feasibility, novelty, and aesthetics ratings of the final design are not related to the amount of time people spend on exploring and editing the overall image versus specific parts of the image. This means that it does not matter how much time is spent on global versus local editing to generate a highly-rated design.

Secondly, the prompts in the global editing mode tend to to be longer, multi-criteria, and more heavily goal oriented than those in the local editing mode. These findings make sense based on the nature of the global editing mode, which is to edit the entire image rather than focusing on specific parts. This seems to lead people to prompt with more words and goals all at once. In the global editing mode, it is found to be beneficial to use more feasibility-oriented prompts and less novelty-related prompts for better feasibility and aesthetics ratings respectively. Also, it is good to keep using multi-criteria prompts over mono-criteria prompts as this helps to reach a higher feasibility rating.

In contrast, the prompts in the local editing mode tend to to be short and single goal oriented (often feasibility rather than novelty or aesthetics). Interestingly, the correlation results show that it is good to use more aesthetics-oriented prompts in the local editing mode as this increases the feasibility rating of the final design. Consistent with this result, close relationship between feasibility and aesthetics is discovered throughout the results. Specifically, the percentage of feasibility-oriented prompts demonstrate a positive correlation to the percentage of aesthetics-oriented prompts. This implies that people intuitively combine prompts that improve both feasibility and aesthetics. Furthermore, though not statistically significant, aesthetics-oriented prompts lead to designs that not only satisfy the raters' visual preferences, but also are rated feasible. This synergistic relationship between aesthetics and feasibility is often referred to as the "aesthetic-usability" or "aesthetic-utility" effect \citep{bib12, bib13}, which underscores the phenomenon where people perceive aesthetically pleasing designs as more usable and effective, even when the functionality remains unchanged.

Finally, the results in this work do not demonstrate a clear way to increase novelty ratings of the final product designs. This suggests that increasing novelty may require other prompting strategies besides manipulating the number and type of goal orientations. For example, Ma, et al. suggest leveraging few-shot learning \citep{ma2023conceptual}. 

In summary, when using text-to-image GenAI tools for product design, it is critical to pay attention to the number and type of goals that are targeted in the prompts perhaps more than to the time spent editing globally and locally and the length of prompts. While longer prompts may seem comprehensive and therefore always effective, they do not consistently lead to better feasibility, novelty, and aesthetic results. Therefore, rather than simply aiming for long prompts, users are advised to strategize to employ specifically oriented prompts to guide DSE more effectively. Incorporating targeted text within prompts can be advantageous in achieving the desired outcomes. What type of targeted prompts should be used then? Initially in the global editing mode, multi-criteria prompts that are both feasibility and aesthetics-oriented are suggested to be used, rather than mono-criteria or novelty-oriented prompts. Then in the local editing mode, it is good to use many aesthetics-oriented prompts. To highlight, aesthetics-oriented prompts are recommended to be employed throughout the DSE process as they benefit the overall rating of the design, especially the feasibility rating.

\subsection{Limitations and Future Work}
There are several limitations in this work. First, it is important to clarify the scope of these findings that it only applies in the context of using GenAI for DSE. If the design is intended for physical manufacturing and beyond, it is crucial to consider additional tools for 3D modeling or physics-based evaluations, and the results from this work do not directly apply. 

Another limitation is that the results of this study, especially those related to time, depend on the usability of the GenAI tool. For example, if the tool's functionalities in the local editing mode are difficult to navigate around, users may spend much more time in the local editing mode than if the functionalities are easier to use. 

Furthermore, the potential impact of user experience and background in design on the outcomes is not directly studied. Incorporating user profiles, such as experience level in design or familiarity with AI tools, could provide deeper insights into how different users might benefit from different strategies.

In addition, some individual biases may be present in the bike ratings. We chose to collect the ratings from 10 raters and only utilize the aggregate values for the analysis. While this method resolves the individual biases more than other methods, such as single expert ratings, the potential discrepancies in the raters like their understanding of feasibility, novelty, and aesthetics may affect the results. Exploring the rater differences in future works can further clarify the results and insights in this work.

Lastly, the study's reliance on a small sample size and a single platform (Leonardo.AI) and text-to-image model (Stable Diffusion) limits the generalizability of its findings. Depending on the performance of the model, especially the training dataset, the nuances of the text prompts may change the generated images. However, the core concept of this paper about using text-to-image models for global and local editing during DSE is generalizable as this is a fundamental tradeoff in design. Therefore, the DSE techniques suggested in this work may be helpful across different models though they may be adopted with caution. Furthermore, the insights in this work about users' DSE behavior using GenAI beyond the specific strategies could have longer lasting value since human cognitive abilities are relatively stable compared to the dynamic computational tools. To further explore the generalizability, future studies with a larger participant pool and other text-to-image models can bolster the findings of this work.  

Future research could also explore the following questions:
1) How does a designer's experience level influence their interaction with GenAI tools? 2) How do different GenAI tools compare in aiding product design exploration?, and 3) How do users' strategies evolve over time with prolonged use of GenAI tools?

\section{Conclusions}

This work investigates how DSE can be conducted effectively when working with GenAI tools, which are often not catered to product design applications. Given the increasing popularity of these tools, many product designers will likely attempt to use these tools in their DSE process. Given the findings in this work, they are likely to be able to  more successfully achieve their design goals, particularly feasibility, novelty, and aesthetics, by strategically guiding their interactions with the tool based on this work rather than using it merely based on their intuition.

The findings in this work provide valuable insights into the user approaches and strategies for DSE success when using text-to-image GenAI tools. By considering these insights and recommendations, researchers and designers can enhance the effectiveness and usability of such tools, and users can utilize these strategies to pursue better product design outcomes. For instance, researchers of Leonardo.AI can study the accurate relationship between feasibility-oriented prompts and the feasibility of the output design during the global editing mode. Additionally, designers can more effectively utilize GenAI for product design, using multi-criteria, feasibility and novelty-oriented prompts in early stages, then mono-criteria, aesthetics-oriented prompts in the later stages. 

\section{Acknowledgements}
Prof. Ahmed and Prof. Lykourentzou extend their gratitude to the MIT MISTI Netherlands Program for their funding support.

\section{Financial Support}

\end{document}